\documentstyle[epsfig,aps,prc]{revtex}

\begin{document}
\title{ One-body density matrix and momentum distribution 
in $s$-$p$ and $s$-$d$ shell nuclei}

\author{Ch. C. Moustakidis and S. E. Massen}
\address{
Department of  Theoretical Physics,
Aristotle University of Thessaloniki,
GR-54006 Thessaloniki, Greece}
%}
%\date{}
\maketitle

\begin{abstract}
Analytical expressions of the one- and two- body terms in the cluster
expansion of the one-body density matrix and momentum distribution of 
the $s$-$p$ and $s$-$d$ shell nuclei with $N=Z$ are derived. They depend 
on the harmonic oscillator parameter $b$ and the parameter $\beta$ which 
originates from the Jastrow correlation function. These parameters have been
determined by least squares fit to the experimental charge form factors. 
The inclusion of short-range correlations increases the high momentum 
component  of the momentum distribution, $n({\bf k})$ for all nuclei we 
have considered while there is an $A$ dependence of $n({\bf k})$ both at 
small values of $k$ and the high momentum component. 
The $A$ dependence of the high momentum component of $n({\bf k})$ becomes 
quite small when the nuclei $^{24}$Mg, $^{28}$Si and $^{32}$S are treated 
as $1d$-$2s$ shell nuclei having the occupation probability of the 
$2s$-state as an extra free parameter in the fit to the form factors.
\end{abstract}

%\newpage
%%%%%%%%%%%%%%%%%%%%%%%%%%%%%%%%%%%%%%%%%%%%%%%%%%%%%%%%
%%%%%%%%%%%%%%%%%%%
\section{INTRODUCTION}
The momentum distribution (MD) is of interest in many research subjects
of modern physics, including those referring to helium, electronic, nuclear, 
and quark systems \cite{Ristig82,Antonov,Silver}.
In the last two decades, there has been significant effort for the
determination of the MD in nuclear matter and  finite nucleon systems 
\cite{Zabolitzky78,Antonov80,Bohigas80,DalRi82,Flynn84,Pandh84,Fantoni84,%
Traini85,Jaminon8586,Benhar86,Schiavilla86,Casas87,Stringari90,%
Stoitsov93,Muther95,Gaidarov95,Ypsilantis95,Antonov96,Arias97,Co92}.
MD is related to the cross sections of various kinds of nuclear reactions.
Specifically, the interaction of particles with nuclei at high energies,
such as  (p,2p), (e,e$'$p), and (e,e$'$) reactions, the nuclear
photo-effect, meson absorption by nuclei, the inclusive proton production
in proton-nucleus collisions, and even phenomena at low energies
such as giant multipole resonances, give significant information about
the nucleon MD.
The experimental evidence obtained from inclusive and exclusive
electron scattering on nuclei established the existence of a high-momentum
component for momenta $k>2\ {\rm fm}^{-1}$
\cite {Day87,Ji90,Ciofi89,Ciofi91}.
It has been shown that, in principle,
mean field theories can not describe  correctly
MD and density distribution  simultaneously \cite{Jaminon8586}
and the main features of MD depend little on the effective mean field
considered \cite{Casas87}.
The reason is that MD is sensitive to short-range and tensor nucleon-nucleon
correlations which are  not included in the mean field theories. Thus,
theoretical approaches, which take into account short range correlations 
(SRC) due to the character of the nucleon-nucleon forces at small
distances, are necessary to be developed.

Zabolitzky and Ey \cite{Zabolitzky78}, employing the coupled-cluster
(or $\exp(S)$) method for the microscopic evaluation of nuclear MD for
the ground states of $^{4}$He and $^{16}$O and using various realistic 
NN-potentials, showed that the contribution of correlations dominates for 
momenta beyond $2\ {\rm fm}^{-1}$.
A realistic interaction  and a many-body approach have been used by
Benhar et al \cite{Benhar86} for the evaluation of MD of $^{12}$C,
$^{16}$O and $^{40}$Ca. Their results have yielded a much larger content
of the high momentum component with respect to the results obtained within 
the Hartree-Fock approach or within methods which take into account 
the effect of correlations phenomenologically.

Bohigas and Stringari \cite{Bohigas80} and Dal Ri et al \cite{DalRi82} 
evaluated the effect of SRC's on the one- and two-  body densities  by 
developing a low order approximation (LOA) in the framework of Jastrow 
formalism. They showed that one-body quantities provide an adequate test 
for the presence of SRC's in nuclei, which indicates that the 
independent-particle wave functions cannot reproduce simultaneously the 
form factor and the MD of a correlated system and also the effect of SRC's 
strongly modify the MD by introducing an important contribution in the 
region $k>2$ fm$^{-1}$.
Stoitsov et al \cite{Stoitsov93} generalised the model of Jastrow
correlations within the LOA of Ref. \cite{Gaudin71}, to heavier nuclei as
$^{16}$O, $^{36}$Ar, $^{40}$Ca. Their analytical expressions for the MD 
show the high momentum tail. They found that there is an A dependence of MD 
for small values 
of $k$, while for large values of $k$ the slope of $\log n(k)$ versus $k$ 
is roughly the same for the above three nuclei as well as for $^4$He. 
The same behaviour of the MD of protons and neutrons for the $A=3,4$ 
nuclei has been found earlier by Schiavilla et al \cite{Schiavilla86}
performing  variational calculations with realistic interactions.
MD for the nuclei $^{4}$He, $^{16}$O and $^{40}$Ca was also calculated by 
Traini and Orlandini \cite{Traini85} within a phenomenological model in 
which dynamical short-range and tensor correlations effects were included.
They showed that SRC increase the high momentum component considerably
while the tensor correlations do not affect the MD appreciably
\cite{Traini85,Dellagiacoma83}.
In heavy nuclei, the local density approximation was used \cite{Stringari90}
for the study of the effect of SRC's in MD and the predictions were in 
agreement with the results of microscopic calculations in nuclear matter 
and in light nuclei.

The influence of SRC's on the MD of nucleons in nuclei has also been
evaluated by M\"{u}ther et al \cite{Muther95} within the Green-function 
approach assuming a realistic meson-exchange potential for the 
nucleon-nucleon interaction. Their analysis on $^{16}$O demonstrates that 
a non-negligible contribution to the  MD should be found in partial waves 
which are unoccupied in the simple shell model.
Another approach is to consider the average occupancy of the relevant
shell-model orbitals \cite{Pandh84}. It has been found that the depletion
of such orbitals can be of the order $15\%$ or more for single-particles
(SP) states below the Fermi energy \cite{Quint87}.

In the various approaches, the MD of the closed shell nuclei $^{4}$He, 
$^{16}$O and $^{40}$Ca as well as of $^{208}$Pb and nuclear matter is 
usually studied. There is no systematic study of the one body density 
matrix (OBDM) and MD which include both the case of closed and open shell 
nuclei. This would be helpful in the calculations of the overlap integrals
and  reactions in that region of nuclei if one wants to go beyond the 
mean field theories \cite{Gaidarov99}.
For that reason, in the present work, we attempt to find some general
expressions for the  OBDM $\rho ({\bf r},{\bf r}')$ and MD $n({\bf k})$ 
which could be used both for closed and open shell nuclei.
This work is a continuation of our previous study \cite{Massen99} on the 
form factors and densities of the $s$-$p$ and $s$-$d$ shell nuclei.
The expression of $\rho ({\bf r},{\bf r}')$ was found, first, using the 
factor cluster expansion of Clark and co-workers
\cite{Clark70,Ristig-Clark,Clark79}
and Jastrow correlation function which introduces SRC for closed shell 
nuclei and then was extrapolated to the case of $N=Z$ open shell nuclei.
$n({\bf k})$ was found by Fourier transform of  $\rho ({\bf r},{\bf r}')$.
These expressions are functionals of the harmonic oscillator (HO) orbitals
and depend on the HO parameter $b$ and the correlation parameter $\beta$.
The values of the parameters $b$ and $\beta$, which we have used for the 
closed shell nuclei $^4$He, $^{16}$O and $^{40}$Ca, are the ones which have 
been determined in Ref. \cite{Massen99} by fit of the theoretical
$F_{ch}(q)$, derived with the same cluster expansion, to the experimental one.
For the open shell nuclei $^{12}$C, $^{24}$Mg, $^{28}$Si and $^{32}$S we 
provide new values for these parameters, which have been found to give a 
better fit to the experimental form factors than in our previous analysis
\cite{Massen99}.
It is found that the high-momentum tail of the MD of all the nuclei we have
considered appears for $k > 2\ {\rm fm}^{-1}$ and also there is an A
dependence of the values of $n(k)$ for
$2\ {\rm fm}^{-1} < k < 5\ {\rm fm}^{-1}$. 
This $A$ dependence of MD was first investigated considering $^{24}$Mg, 
$^{28}$Si and $^{32}$S as $1d$ shell nuclei. Next we treated the above 
nuclei as $1d$-$2s$ shell nuclei having the occupation probability of the 
$2s$ state as an extra free parameter in the fit of the form factors. 
The $A$ dependence is quite small in the second case.

The paper is organised as follows. In Sec. II the general expressions
of the correlated OBDM and MD are derived using
a Jastrow correlation function. In Sec. III the analytical expressions
of the above quantities for the $s$-$p$ and $s$-$d$ shell nuclei,
in the case of the HO orbitals, are given. Numerical results are reported
and discussed in Sec. IV, while the summary of the present work is
given in Sec. V.
%%%%%%%%%%%%%%%%%%%%%%%%%%%%%%%%%%%%%%%%%%%%%%%%%%%%%%%%%%%%%%%%%%%%%%%%

\section{CORRELATED ONE-BODY DENSITY MATRIX AND MOMENTUM DISTRIBUTION}

A nucleus with $A$ nucleons is described by the wave function
$\Psi({\bf r}_1,{\bf r}_2,...,{\bf r}_A) $ which depends on $3A$
coordinates as well as on  spins and isospins.
The evaluation of the single particle characteristics
of the system needs the one-body density matrix \cite{Dirac30,Lowdin55}
\begin{equation}
\rho ({\bf r},{\bf r}')= \int \Psi^{*}({\bf r},{\bf r}_2,...,{\bf r}_A) \
\Psi ({\bf r}',{\bf r}_2,\cdots,  {\bf r}_A) \ {\rm d}{\bf r}_2 \cdots
{\rm d}{\bf r}_A \ ,
\end{equation}
where the integration is carried out over the radius vectors
${\bf r}_2,\cdots, {\bf r}_A$ and summation
over spin and isospin variables is implied. $\rho ({\bf r},{\bf r}')$ can
also be represented by the form
\begin{equation}
\rho ({\bf r},{\bf r}')=\frac{\langle \Psi|{\bf O}_{\bf r \bf r'}
|\Psi '\rangle}{\langle\Psi|\Psi\rangle}\
 =\ N \langle \Psi|{\bf O}_{\bf r \bf r'}|\Psi '\rangle
=N  \langle {\bf O}_{\bf r \bf r'} \rangle \ ,
\label{den-rr2}
\end{equation}
where $\Psi '=\Psi({\bf r}_1',{\bf r}_2',...,{\bf r}_A') $
and $N$ is the normalization factor.
The one-body "density operator" ${\bf O}_{{\bf r r}'}$,
has the form
\begin{equation}
{\bf O}_{{\bf r r}'} =\sum_{i=1}^{A} \ \delta({\bf r}_i-{\bf r})
\delta({\bf r}_i'-{\bf r}')\prod_{j \neq i}^{A}\delta({\bf r}_j-{\bf r}_j')
\ .
\label{O(rr)}
\end{equation}

In the case where the nuclear wave function
$\Psi$ can be expressed as a Slater determinant depending on the
SP wave functions $\phi_i({\bf r})$ we have
\begin{equation}
\rho_{SD}({\bf r},{\bf r}')=\sum_{i=1}^{A}\phi_i^{*}({\bf r}) \phi_i({\bf r}')
\end{equation}

The diagonal elements of the OBDM give
the density distribution
\begin{equation}
\rho({\bf r},{\bf r})=\rho({\bf r}) \ ,
\end{equation}
while the MD is given by the Fourier transform of $\rho({\bf r},{\bf r}')$,
\begin{equation}
n({\bf k})=\frac{1}{(2\pi)^3}\int \exp[i{\bf k}({\bf r}-{\bf r}')] \
\rho({\bf r},{\bf r}') \ {\rm d}{\bf r} \ {\rm d}{\bf r}' \ .
\end{equation}
In the case of a Slater determinant, MD takes the form
\begin{equation}
n_{SD}({\bf k})=
\sum_{i=1}^{A}\tilde{\phi}_i^{*}({\bf k}) \tilde{\phi}_i({\bf k}) \ ,
\end{equation}
where
\begin{equation}
\tilde{\phi}_{i}({\bf k})=\frac{1}{(2\pi)^{3/2}}
\int \phi_{i}({\bf r}) \exp[i{\bf k}{\bf r}] {\rm d}{\bf r}.
\end{equation}

The second moment of the MD is related to the
expectation value of the kinetic energy, $\langle {\bf T}\rangle$,
by the expression
\begin{equation}
\langle {\bf T}\rangle=\frac{\hbar^2}{2m} \int n({\bf k})k^2 {\rm d}
{\bf k} .
\end{equation}

%%%%%%%%%%%%%%%%%%%%%%%%
\subsection{One-body density matrix \label{sub2.2}}

If we denote the model operator, which introduces SRC, by
$\cal{F}$,
an eigenstate $\Phi$ of the model system corresponds to an eigenstate
\begin{equation}
\Psi={\cal F}\Phi
\label{eq1}
\end{equation}
of the true system.

Several restrictions can be made on the model operator $\cal{F}$,
as for example, that it depends on (the spins, isospins and) relative
coordinates and momenta of the particles in the system, that be a scalar
with respect to rotations etc. \cite{Brink67}. Further, it is required that
$\cal{F}$ be translationally invariant and symmetrical in its arguments
$1 \cdots i \cdots A$ and possesses the cluster property. That is if any
subset $i_1 \cdots i_p$ of the particles is removed far from the rest
$i_{p+1} \cdots i_A$, $\cal{F}$ decomposes into a product of two factors,
${\cal F}(1\cdots A) =
{\cal F}(i_1 \cdots i_p) \ {\cal F}(i_{p+1} \cdots i_A)$ \cite{Clark79}.
In the present work $\cal{F}$ is taken to be of the Jastrow type
\cite{Jastrow55},
\begin{equation}
{\cal F}=\prod_{i<j}^{A}f(r_{ij})\ ,
\end{equation}
where $f(r_{ij})$ is the state-independent correlation function
of the form
\begin{equation}
f(r_{ij})=1-\exp[-\beta({\bf r}_i-{\bf r}_j)^2] \ .
\label{fr-ij}
\end{equation}

The correlation function $f(r_{ij})$ goes to 1 for large values of
$r_{ij} = \mid {\bf r}_i - {\bf r}_j \mid$ and it goes to 0 for
$r_{ij} \rightarrow 0$. It is obvious that the effect of SRC, introduced
by the function $f(r_{ij})$, becomes large when the SRC parameter
$\beta$ becomes small and vice versa.

In order to evaluate the correlated one-body density matrix
$\rho_{cor}({\bf r},{\bf r}')$,
we consider, first, the generalized integral
\begin{equation}
I(\alpha)=\langle\Psi | \exp [\alpha  I(0) 
{\bf O}_{{\bf rr}'}] |\Psi'\rangle \ ,
\label{I-1}
\end{equation}
corresponding to the one-body "density operator" ${\bf O}_{{\bf rr}'}$ 
(given by (\ref{O(rr)})), from which we have
\begin{equation}
\langle{\bf O}_{{\bf rr}'}\rangle = \left[ \frac{\partial \ln I (\alpha)}
{\partial \alpha} \right]_{\alpha=0} .
\label{I-2}
\end{equation}
For the cluster analysis of equation (\ref{I-2}), we consider the sub-product
integrals \cite{Clark70,Ristig-Clark,Clark79}, for the sub-systems of 
the $A$-nucleons system
\begin{eqnarray}
I_i(\alpha)&=&\langle i\mid {\cal F}^{\dag}(r_{1})\exp[\alpha I_i(0)
{\bf o}_{\bf{rr}'}(1)]
{\cal F}(r_{1}')\mid i'\rangle \ , \nonumber\\
I_{ij}(\alpha)&=&\langle ij\mid {\cal F}^{\dag}(r_{12})
\exp[\alpha I_{ij}(0) {\bf o_{rr'}}(2)]
{\cal F}(r_{12}')\mid i'j'\rangle_{\rm a}  \ , \nonumber\\
I_{ijk}(\alpha)&=&\langle ijk\mid
{\cal F}^{\dag}(r_{12}){\cal F}^{\dag}(r_{13}){\cal F}^{\dag}(r_{23})
\exp[\alpha I_{ijk}(0){\bf o_{rr'}}(3)]
{\cal F}(r_{12}'){\cal F}(r_{13}'){\cal F}(r_{23}')
\mid i'j'k' \rangle_{\rm a} \ ,
\nonumber\\
&.&\nonumber\\
&.&\nonumber\\
&.&\nonumber\\
I_{12 \cdots A}&=&I(\alpha) \ ,
\end{eqnarray}
where the operators ${\bf o}_{\bf{rr}'}(1)$, ${\bf o}_{\bf{rr}'}(2)$, $\ldots$
have the form
\begin{eqnarray}
{\bf o_{rr'}}(1)&=&\delta({\bf r}_1-{\bf r})\delta({\bf r}_1'-{\bf r}') \ ,
\nonumber\\
{\bf o_{rr'}}(2)&=&\delta({\bf r}_1-{\bf r})\delta({\bf r}_1'-{\bf r}')
\delta({\bf r}_2-{\bf r}_2')+
\delta({\bf r}_2-{\bf r})\delta({\bf r}_2'-{\bf r}')
\delta({\bf r}_1-{\bf r}_1') \ ,
\end{eqnarray}
and so on.The factor cluster decomposition of the above integrals, 
following the factor cluster expansion of Ristig,Ter Low, and Clark 
\cite{Clark70,Ristig-Clark,Clark79}, gives
\begin{equation}
\langle {\bf O}_{{\bf rr}'}\rangle =
 \langle {\bf O}_{{\bf rr}'}\rangle_1 +
\langle {\bf O}_{{\bf rr}'}\rangle_2 + \cdots +
\langle {\bf O}_{{\bf rr}'}\rangle_A  \ ,
\end{equation}
where
\begin{eqnarray}
\langle {\bf O}_{{\bf rr}'} \rangle _1 &=&
 \sum_{i=1}^{A} \left[ \frac{\partial \ln I_{i}(\alpha)}{\partial \alpha}
\right]_{\alpha=0} \ ,
\end{eqnarray}
\begin{eqnarray}
\langle {\bf O}_{{\bf rr}'} \rangle_2 &=&
\sum_{i<j}^{A} \frac{\partial}{\partial \alpha} \left[
 \ln I_{ij}(\alpha) - \ln I_i(\alpha) -  \ln I_j(\alpha) \right]_{\alpha=0}\ ,
%\label{Or2-1} \ ,
\end{eqnarray}
and so on. ${\cal F}(r_1)$ is chosen to be the identity operator.

Three- and many-body terms will be neglected in the present analysis.
Thus, in the two-body approximation, $\rho_{cor}({\bf r},{\bf r}')$,
defined by Eq. (\ref{den-rr2}), is written
\begin{equation}
\rho_{cor}({\bf r},{\bf r}') \approx N [ \langle
 {\bf O}_{{\bf rr}'} \rangle_1
+ \langle {\bf O}_{{\bf rr}'}\rangle_{22} -
\langle {\bf O}_{{\bf rr}'}\rangle_{21}] \ ,
\end{equation}
where
\begin{equation}
\langle {\bf O}_{{\bf rr}'} \rangle _1 =
 \sum_{i=1}^{A} \langle i \mid {\bf o}_{{\bf rr}'}(1) \mid i' \rangle \ ,
\label{Or1-2}
\end{equation}
\begin{equation}
\langle {\bf O}_{{\bf rr}'} \rangle_{22} =
 \sum_{i<j}^{A}\langle ij \mid {\cal F}^{\dag}(r_{12})
{\bf o}_{{\bf rr}'}(2) {\cal F}(r'_{12}) \mid i'j' \rangle_{\rm a},
\end{equation}
\begin{equation}
\langle {\bf O}_{{\bf rr}'} \rangle_{21} =
 \sum_{i<j}^{A} \langle ij \mid {\bf o}_{{\bf rr}'}(2)\mid i'j'
\rangle_{\rm a}
\label{Or2-1}
\end{equation}
If the two-body operator ${\cal F}(r'_{12})$ is taken to be the correlation 
function given by Eq. (\ref{fr-ij}), then
\begin{equation}
{\cal F}^{\dag}(r_{12}){\bf o}_{{\bf rr}'}(2) {\cal F}(r'_{12})=
{\bf o}_{{\bf rr}'}(2) \left[ 1 -  {\rm g}_{1}({\bf r},{\bf r}_2) -
{\rm g}_{2}({\bf r}',{\bf r}_2) +  {\rm g}_{3}({\bf r},{\bf r}',{\bf r_2}) 
\right],
\label{F12F12}
\end{equation}
where
\begin{eqnarray}
{\rm g}_1({\bf r},{\bf r}_2)& =& \exp [-\beta(r^2 + r_2^2)]
\exp [2\beta {\bf r r}_2 ], \quad
{\rm g}_2({\bf r}',{\bf r}_2) = {\rm g}_1({\bf r}',{\bf r}_2),
\nonumber\\
{\rm g}_3({\bf r},{\bf r}',{\bf r}_2)& = &\exp [-\beta (r^2+{r'}^2)]
\exp [-2\beta r_2^2]\exp [2\beta ({\bf r}+ {\bf r}'){\bf r}_2 ] 
\end{eqnarray}
and the term $\langle {\bf O}_{\bf rr'} \rangle_{22}$ is written
\begin{equation}
\langle {\bf O}_{\bf rr'} \rangle_{22} =
\langle {\bf O}_{\bf rr'} \rangle_{21}
- O_{22}({\bf r},{\bf r}',{\rm g}_1) - O_{22}({\bf r},{\bf r}',{\rm g}_2)
+ O_{22}({\bf r},{\bf r}',{\rm g}_3) \ ,
\label{O22-2}
\end{equation}
where
\begin{equation}
O_{22}({\bf r},{\bf r}',{\rm g}_{\ell}) =
 \sum_{i<j}^{A}\langle ij \mid {\bf o}_{{\bf rr}'}(2)
{\rm g}_\ell({\bf r},{\bf r}',{\bf r}_2)\mid i'j' \rangle_{\rm a} \ ,
\quad \ell=1,2,3 \ .
\end{equation}

Thus, $\rho_{cor}({\bf r},{\bf r}')$ takes the form
\begin{equation}
\rho_{cor}({\bf r},{\bf r}') \approx
N [ \langle {\bf O}_{{\bf rr}'} \rangle_1
- O_{22}({\bf r},{\bf r}',{\rm g}_1) - O_{22}({\bf r},{\bf r}',{\rm g}_2)
+ O_{22}({\bf r},{\bf r}',{\rm g}_3) ] \ .
\label{Dp-3}
\end{equation}
%%%%

This is also expressed in the following form
\begin{eqnarray}
\rho_{cor}({\bf r},{\bf r}')& \approx& N \left[ \rho_{SD}({\bf r},{\bf r}')
+\int \left[
-{\rm g}_1({\bf r},{\bf r}_2)
-{\rm g}_2({\bf r}',{\bf r}_2)
+{\rm g}_3({\bf r},{\bf r}',{\bf r}_2) \right] \right. \nonumber\\
\nonumber\\
& &\times \left. \left[\rho_{SD}({\bf r},{\bf r}')
\rho_{SD}({\bf r}_2,{\bf r}_2)- \rho_{SD}({\bf r},{\bf r}_2)
\rho_{SD}({\bf r}_2,{\bf r}')\right]
{\rm d}{\bf r}_2   \frac{}{}\right],
\label{Eq-ro}
\end{eqnarray}
where $\rho_{SD}({\bf r},{\bf r'})$ is the uncorrelated OBDM
associated with the Slater determinant.

It should be noted that a similar expression for
$\rho_{cor}({\bf r},{\bf r}')$, given by
Eq. (\ref{Eq-ro}), was derived by Gaudin et al. \cite{Gaudin71}
in the framework  of LOA.
Their expansion
contains one- and two-body terms and a part of the three-body term
which was chosen so that the normalization of the wave function was
preserved. Expression (\ref{Eq-ro}) of the present work has only one- and
two-body terms and the normalization of the wave function is preserved
by the normalization factor $N$.

In the above expression of $\rho_{cor}({\bf r},{\bf r}')$, the one-body
contribution to the OBDM is well known and is given by the equation
\begin{equation}
\langle {\bf O}_{{\bf rr}'} \rangle_1=\rho_{SD}({\bf r},{\bf r}')=
 \frac{1}{\pi} \sum_{nl} \eta_{nl} (2l+1)
 \phi^{*}_{nl}(r) \phi_{nl}(r') P_l(\cos \omega_{rr'} )
\label{O1-3}
\end{equation}
where $\eta_{nl}$ are the occupation probabilities of the states $nl$
(0 or 1 in the case of closed shell nuclei) and $\phi_{nl}(r)$ is the 
radial part of the SP wave function and  $\omega_{rr'}$ the angle between 
the vectors ${\bf r}$ and ${\bf r}'$.

The term $O_{22}({\bf r},{\bf r}',{\rm g}_\ell)$, performing the 
spin-isospin summation and the angular integration, takes the general form
\begin{eqnarray}
O_{22}({\bf r},{\bf r}',{\rm g}_{\ell})& = & 4 \sum_{n_i l_i,n_j l_j} 
\eta_{n_i l_i} \eta_{n_j l_j} (2 l_i +1) (2 l_j +1 )  \nonumber \\
&  &\times \left[ 4 A_{n_il_in_jl_j}^{n_il_i n_jl_j,0 }
({\bf r},{\bf r}',{\rm g}_\ell) - \sum_{k=0}^{l_i +l_j}
\langle l_i 0 l_j 0 \mid k0 \rangle^2
 A_{n_il_in_jl_j}^{n_jl_j n_il_i,k}({\bf r},{\bf r}',{\rm g}_\ell) \right],
\ \ell=1, 2, 3  , 
\label{O22-g-3}
\end{eqnarray}
where
\begin{eqnarray}
A_{n_1l_1n_2l_2}^{n_3l_3n_4l_4,k}({\bf r},{\bf r}',{\rm g}_1)& =&
\frac{1}{4\pi}\phi^{*}_{n_1l_1}(r) \
\phi_{n_3l_3}(r') \ \exp[-\beta r^2] \ P_{l_3}(\cos\omega_{rr'})
 \nonumber \\
& &\times \int_{0}^{\infty}\phi^{*}_{n_2l_2}(r_2) \ \phi_{n_4l_4}(r_2) \
\exp[-\beta r_{2}^2] \ i_k (2 \beta r r_2)
\ r_{2}^{2} \ {\rm d}r_2  ,
\label{A-O22-1}
\end{eqnarray}
and the matrix element
$A_{n_1l_1n_2l_2}^{n_3l_3n_4l_4,k}({\bf r},{\bf r}',{\rm g}_2)$
can be found from (\ref{A-O22-1}) replacing
${\bf r}\longleftrightarrow {\bf r}'$ and  $n_1l_1\longleftrightarrow n_3l_3$
while the matrix element corresponding to the factor ${\rm g}_3$ is
\begin{eqnarray}
A_{n_1l_1n_2l_2}^{n_3l_3n_4l_4,k}({\bf r},{\bf r}',{\rm g}_3)& =&
\frac{1}{4\pi}\phi^{*}_{n_1l_1}(r) \
\phi_{n_3l_3}(r') \ \exp[-\beta (r^2+r'^2)] \
\Omega_{l_1l_3}^{k}(\omega_{rr'})
\times \nonumber \\
& & \int_{0}^{\infty}\phi^{*}_{n_2l_2}(r_2) \ \phi_{n_4l_4}(r_2) \
\exp[- 2 \beta r_{2}^2] \ i_k (2 \beta |{\bf r}+{\bf r}'| r_2) \
r_{2}^{2} \  {\rm d}r_2 ,
\label{A-O22-3}
\end{eqnarray}
In Eqs. (\ref{A-O22-1}) and (\ref{A-O22-3})  the modified spherical Bessel
function, $i_k(z)$, comes from the expansion of the exponential function
$\exp[2\beta {\bf x}_1 {\bf x}_2]$ of the factors ${\rm g}_{\ell}$
in spherical harmonics, that is
\[
\exp[2\beta {\bf x}_1 {\bf x}_2] = 4 \pi \sum_{km_k}
i_k(2\beta x_1 x_2) Y^{*}_{km_k}(\Omega_1)Y_{km_k}(\Omega_2),
\]
while the factor $\Omega_{l_1l_3}^{k}(\omega_{rr'})$ which depends on
the directions of ${\bf r}$ and ${\bf r}'$ is,
\begin{equation}
\Omega_{l_1l_3}^{k}(\omega_{rr'})=
\sum_{m_1,m_3}
\frac{\langle l_1 -m_1 l_3 m_3 \mid k 0 \rangle}
{\langle l_1 0 l_3 0 \mid k 0 \rangle}
\left( \frac{(l_1-m_1)!}{(l_1+m_1)!}\frac{(l_3-m_3)!}{(l_3+m_3)!}
\right)^{\frac{1}{2}} \ P_{l_1}^{m_1}(\cos\omega_r)
\ P_{l_3}^{m_3}(\cos\omega_{r'}) \ ,
\label{omega-1}
\end{equation}
where $\omega_r$ and $\omega_{r'}$ are the angles between the vectors
${\bf r},\ {\bf r}+{\bf r}'$ and ${\bf r}',\ {\bf r}+{\bf r}'$,
respectively. The final expression of $\Omega_{l_1l_3}^{k}(\omega_{rr'})$
depends on $\omega_{rr'}$, the angle  between the vectors
${\bf r}$ and ${\bf r}'$.

The expression of the term $O_{22}({\bf r},{\bf r}', {\rm g}_\ell)$
depends on the SP
wave functions and so it is suitable to be used for analytical
calculations with the HO orbitals and in principle for numerical
calculations with more realistic SP orbitals.
Expressions (\ref{O1-3}) and (\ref{O22-g-3}) were derived for the closed
shell nuclei with $N=Z$, where $\eta_{nl}$ is 0 or 1. For the open shell
nuclei (with $N=Z$) we use the same expressions, where now
$0 \le \eta_{nl} \le 1$. In this way the mass dependence of the
correlation parameter $\beta$ and the OBDM or MD can be studied.

Finally, using the known values of the Clebsch-Gordan coefficients,
Eq. (\ref{O22-g-3}), for the case of $s$-$p$ and $s$-$d$ shell nuclei,
takes the form
\begin{eqnarray}
& &O_{22}({\bf r},{\bf r}',{\rm g}_\ell)= \nonumber \\
& &4\left[ \frac{}{}
3A_{0000}^{0000,0}({\bf r},{\bf r}',{\rm g}_{\ell})\eta_{1s}^{2}
+\left[33 A_{0101}^{0101,0}({\bf r},{\bf r}',{\rm g}_{\ell}) -
6 A_{0101}^{0101,2}({\bf r},{\bf r}',{\rm g}_{\ell}) \right] \eta_{1p}^2
+3A_{1010}^{1010,0}({\bf r},{\bf r}',{\rm g}_{\ell})\eta_{2s}^2 \right. 
\nonumber \\
& &\nonumber\\
& &+ \left[ 95 A_{0202}^{0202,0}({\bf r},{\bf r}',{\rm g}_{\ell}) -
\frac{50}{7}A_{0202}^{0202,2}({\bf r},{\bf r}',{\rm g}_{\ell})
-\frac{90}{7}A_{0202}^{0202,4}({\bf r},{\bf r}',{\rm g}_{\ell}) \right] 
\eta_{1d}^2    \nonumber \\
& &\nonumber\\
& &+\left[ 12A_{0001}^{0001,0}({\bf r},{\bf r}',{\rm g}_{\ell})+
12A_{0100}^{0100,0}({\bf r},{\bf r}',{\rm g}_{\ell})
-3A_{0001}^{0100,1}({\bf r},{\bf r}',{\rm g}_{\ell})-
3A_{0100}^{0001,1}({\bf r},{\bf r}',{\rm g}_{\ell})  \right] \eta_{1s} 
\eta_{1p}   \nonumber\\
& &\nonumber\\
& &+ \left[ 20A_{0002}^{0002,0}({\bf r},{\bf r}',{\rm g}_{\ell})+
20A_{0200}^{0200,0}({\bf r},{\bf r}',{\rm g}_{\ell})
-5A_{0002}^{0200,2}({\bf r},{\bf r}',{\rm g}_{\ell})-
5A_{0200}^{0002,2}({\bf r},{\bf r}',{\rm g}_{\ell})  \right] \eta_{1s} 
\eta_{1d}    \nonumber\\
& &\nonumber\\
& &+ \left[ 4A_{0010}^{0010,0}({\bf r},{\bf r}',{\rm g}_{\ell})+
4A_{1000}^{1000,0}({\bf r},{\bf r}',{\rm g}_{\ell})
-A_{0010}^{1000,0}({\bf r},{\bf r}',{\rm g}_{\ell}) -
A_{1000}^{0010,0}({\bf r},{\bf r}',{\rm g}_{\ell}) \right] \eta_{1s} 
\eta_{2s}   \nonumber\\
& &\nonumber\\
& &+ \left[ 60 A_{0102}^{0102,0}({\bf r},{\bf r}',{\rm g}_{\ell}) +
60 A_{0201}^{0201,0}({\bf r},{\bf r}',{\rm g}_{\ell})
-6 A_{0102}^{0201,1}({\bf r},{\bf r}',{\rm g}_{\ell}) -
6 A_{0201}^{0102,1}({\bf r},{\bf r}',{\rm g}_{\ell})  \right.
 \nonumber\\
& &\nonumber\\
& &- \left. 9 A_{0102}^{0201,3}({\bf r},{\bf r}',{\rm g}_{\ell}) -
9 A_{0201}^{0102,3}({\bf r},{\bf r}',{\rm g}_{\ell}) \right]
\eta_{1p} \eta_{1d}    
\nonumber\\
& &\nonumber\\
& &+ \left[ 12A_{0110}^{0110,0}({\bf r},{\bf r}',{\rm g}_{\ell}) +
12A_{1001}^{1001,0}({\bf r},{\bf r}',{\rm g}_{\ell}) 
-3A_{0110}^{1001,1}({\bf r},{\bf r}',{\rm g}_{\ell})-
3A_{1001}^{0110,1}({\bf r},{\bf r}',{\rm g}_{\ell}) \right] \eta_{1p} 
\eta_{2s}   \nonumber\\
& &\nonumber\\
& &\left.+ \left[ 20A_{0210}^{0210,0}({\bf r},{\bf r}',{\rm g}_{\ell})+
20A_{1002}^{1002,0}({\bf r},{\bf r}',{\rm g}_{\ell})
-5A_{0210}^{1002,2}({\bf r},{\bf r}',{\rm g}_{\ell})-
5A_{1002}^{0210,2}({\bf r},{\bf r}',{\rm g}_{\ell})  
 \right] \eta_{_{1d}} \eta_{2s}
\frac{}{} \right]  .
%\nonumber\\
%& &
\label{O22-A}
\end{eqnarray}

It should be noted that Eqs. (\ref{O22-g-3}) and (\ref{O22-A}) are
also valid for the cluster expansion of the density distribution and the
form factor as it has been found in ref. \cite{Massen99} and also
in the cluster expansion of the MD. The only difference
is the expressions of the matrix elements $A$.

\subsection{Momentum distribution \label{sub2.3}}
            %%%%%%%%%%%%%%%%%%%%%%

The MD for the above mentioned nuclei can be
found either by following the same cluster expansion or by taking the Fourier
transform of $\rho({\bf r}, {\bf r}')$ given by (\ref{Dp-3}). In both cases
the correlated momentum distribution takes the form
\begin{equation}
n_{cor}({\bf k}) \approx N \left[ \langle  {\bf \tilde{O}}_{\bf k} \rangle_1
- 2 \tilde{O}_{22}({\bf k},{\rm g}_1) +
\tilde{O}_{22}({\bf k}, {\rm g}_3) \right] \ ,
\label{ncor-1}
\end{equation}
where
\begin{equation}
\langle  {\bf \tilde{O}}_{\bf k} \rangle_1 =n_{SD}({\bf k})=
\sum_{i=1}^{A}\tilde{\phi}_i^*({\bf k})\tilde{\phi}_i({\bf k}).
\label{O1k-1}
\end{equation}

The term $\tilde{O}_{22}({\bf k},{\rm g}_\ell)$, as in the case of OBDM, 
is given again by the right-hand side of Eqs. (\ref{O22-g-3}) and 
(\ref{O22-A}) replacing the matrix elements
$A_{n_1l_1n_2l_2}^{n_3l_3n_4l_4,k}({\bf r},{\bf r}',{\rm g}_{\ell})$, 
defined by Eqs. (\ref{A-O22-1}) and (\ref{A-O22-3}),
by the Fourier transform of them, that is by the matrix elements
%$\tilde{A}_{n_1l_1n_2l_2}^{n_3l_3n_4l_4,k}({\bf k},{\rm g}_{\ell})$,
%
\begin{equation}
\tilde{A}_{n_1l_1n_2l_2}^{n_3l_3n_4l_4,k}({\bf k},{\rm g}_{\ell}) =
\frac{1}{(2\pi)^3} \int 
A_{n_1l_1n_2l_2}^{n_3l_3n_4l_4,k}({\bf r},{\bf r}',{\rm g}_{\ell})
\exp[i{\bf k}({\bf r}-{\bf r}')]  {\rm d}{\bf r} {\rm d}{\bf r}', 
\quad \ell=1,3 \ .
\label{Ak-O22-1}
\end{equation}

As in the case of the OBDM, expression (\ref{ncor-1}) is suitable for 
the study of the MD for the $s$-$p$ and $s$-$d$ shell nuclei and also for 
the study of the mass dependence of the kinetic energy of these nuclei. 
The mean value of the kinetic energy has the form
\begin{equation}
\langle {\bf T} \rangle= 
N [ \langle {\bf T} \rangle_1 - 2 T_{22}({\rm g}_1) + T_{22}({\rm g}_3) ]\ ,
\label{Kin-1}
\end{equation}
where
\begin{equation}
\langle {\bf T} \rangle_1=\frac{\hbar^2}{2m}
\int k^2 n_{SD}({\bf k}) {\rm d}{\bf k}\ ,
\quad T_{22}({\rm g}_{\ell})=\frac{\hbar^2}{2m}
\int k^2 \tilde{O}_{22}({\bf k},{\rm g}_{\ell}) {\rm d}{\bf k}\ ,
\quad \ell =1,3 \ .
\end{equation}
%%%%%%%%%%%%%%%%%%%%%%%%%%%

\section{ANALYTICAL EXPRESSIONS \label{sec-3}}
         %%%%%%%%%%%%%%%%%%%%%%
In the case of the HO wave functions, with radial part in coordinate
and momentum space,
\begin{eqnarray}
\phi_{nl}(r)&=&N_{nl}b^{-3/2}  r_{b}^{l}
L_n^{l+\frac{1}{2}} \left( r_{b}^{2} \right)
{\rm e}^{-r_{b}^{2}/2}\ , \,\, r_b=r/b \ ,  \nonumber\\
\tilde{\phi}_{nl}(k)&=& i^l (-1)^{n+l}  N_{nl} b^{3/2}
k_{b}^{l} L_n^{l+\frac{1}{2}} \left( k_{b}^{2} \right) \
{\rm e}^{- k_{b}^{2}/2} \ , \,\, k_b=kb \ ,
\end{eqnarray}
where
\[
N_{nl}=\left(\frac{2 n!}{\Gamma(n+l+\frac{3}{2})} \right)^{1/2} ,
\]
analytical expressions of the one-body terms, 
$\langle {\bf O_{rr'}}\rangle_1$ and 
$\langle {\bf \tilde{O}}_{\bf k} \rangle_1$ 
as well as of the matrix elements
$A_{n_1l_1n_2l_2}^{n_3l_3n_4l_4,k}({\bf r},{\bf r}',{\rm g}_\ell)$
and $\tilde{A}_{n_1l_1n_2l_2}^{n_3l_3n_4l_4, k}({\bf k},{\rm g}_\ell)$,
which have been defined in Sec. II, can be found.
From these expressions, the analytical expressions of the terms
$O_{22}({\bf r},{\bf r}',{\rm g}_{\ell})$ and
$\tilde{O}_{22}({\bf k},{\rm g}_{\ell})$, defined by Eq.
(\ref{O22-A}), can also be found.

The expressions of the one-body terms, $\langle {\bf O_{rr'}}\rangle_1$ and 
$\langle {\bf \tilde{O}}_{\bf k} \rangle_1$, have the forms
\begin{eqnarray}
\langle {\bf O}_{{\bf rr}'} \rangle_1 =
\rho_{SD}({\bf r},{\bf r}')&=& \frac{2}{\pi^{3/2}b^3}
\left[ \frac{}{}  2\eta_{1s}+3\eta_{2s} -2 \eta_{2s} (r_b^2+ {r_b'}^2)
 + 4\eta_{1p}r_b r_b' \cos\omega_{rr'}       \right.  \nonumber \\
\nonumber \\
& &+  \left.   \frac{4}{3} [ \eta_{2s}+ \eta_{1d}(3\cos^2\omega_{rr'} -1)
r_b^2 {r_b'}^2 ] \frac{}{}  \right] \exp[-(r_b^2+{r_b'}^2)/2]
\end{eqnarray}
\begin{equation}
\langle  {\bf \tilde{O}}_{\bf k} \rangle_1= n_{SD}({\bf k})=
\frac{2b^3}{\pi^{3/2}}
\exp[-k_b^2] \
\sum_{k=0}^{2}C_{2k} k_b^{2k} \ ,
\end{equation}
where  the coefficients  $C_{2k}$ are
\begin{equation}
C_0 = 2\eta_{1s}+3\eta_{2s} \ , \quad C_2 =  4(\eta_{1p}-\eta_{2s}) \ ,
\quad C_4 =  \frac{4}{3} (2 \eta_{1d}+  \eta_{2s}) \ .
\end{equation}

%%%%
The analytical expressions of the matrix element
$A_{n_1l_1n_2l_2}^{n_3l_3n_4l_4,k}({\bf r}, {\bf r}',{\rm g}_{\ell})$
$(\ell=1,3)$ have the form
%which are given by Eqs. (\ref{A-O22-1}) and (\ref{A-O22-3})
%
\begin{eqnarray}
A_{n_1l_1n_2l_2}^{n_3l_3n_4l_4,k}({\bf r}, {\bf r}',{\rm g}_1)&=&
B_0 \ b^{-3} y^k \  r_b^{k+l_1}  {r'_b}^{l_3}
\ L_{n_1}^{l_1+\frac{1}{2}}(r_b^2) \ L_{n_3}^{l_3+\frac{1}{2}}({r'_b}^2)
\exp \left[- \frac{1+3y}{2(1+y)}r_b^2-\frac{1}{2}{r'_b}^2 \right]
\nonumber\\
& &
\times  P_{l_{3}}(\cos\omega_{rr'}) \sum_{w_{2}=0}^{n_2}\sum_{w_{4}=0}^{n_4}
B_{w_2 w_4}(y)\ L_{\frac{1}{2}(l_2+l_4-k)+w_2+w_4}^{k+\frac{1}{2}}
 \left(\frac{-y^2 }{1+y} r_b^2  \right)\ ,
\label{Ar-an-g1}
\end{eqnarray}
and
\begin{eqnarray}
A_{n_1l_1n_2l_2}^{n_3l_3n_4l_4,k}({\bf r}, {\bf r}',{\rm g}_3)&=&
B_0 \ b^{-3} y^k  \mid {\bf r}_b+ {\bf r}'_b \mid ^k  r_b^{l_1}  {r'_b}^{l_3}
\ L_{n_1}^{l_1+\frac{1}{2}}(r_b^2) \
L_{n_3}^{l_3+\frac{1}{2}}({r'_b}^2) \Omega_{l_1 l_3}^k(\omega_{rr'})
 \nonumber\\
& &\times \exp \left[-\frac{1+2y}{2}(r_b^2+{r'_b}^2) \right]
\exp \left[\frac{y^2}{1+2y}({\bf r}_b +{\bf r}'_b)^2 \right]
\nonumber\\
& &\times \sum_{w_{2}=0}^{n_2}\sum_{w_{4}=0}^{n_4}
B_{w_2 w_4}(2y) \
L_{\frac{1}{2}(l_2+l_4-k)+w_2+w_4}^{k+\frac{1}{2}}
 \left(-\frac{y^2}{1+2y}({\bf r}_b+{\bf r}'_b)^2 \right)
\label{Ar-an-g3}
\end{eqnarray}
where $y=\beta b^2$ and
\begin{equation}
B_0 = \frac{1}{16  \sqrt{\pi} } \left( \prod_{i=1}^{4}  N_{n_il_i} \right) ,
\end{equation}
\begin{equation}
B_{w_2 w_4}(z)= [ \frac{1}{2}(l_2+l_4-k) + w_2+w_4 ]!
 \prod_{i=2,4} \frac{(-1)^{w_i}}{w_i!}
\left( \!\!  \begin{array}{c}
n_{i}+l_{i}+\frac{1}{2} \\
n_{i}-w_i
\end{array} \!\! \right)
(1+z)^{-\frac{1}{2}l_i -w_i -\frac{1}{2}(3 +k)}  \ ,
\label{B-w2w4}
\end{equation}
while the one corresponding to the factor ${\rm g}_2$ can be found from 
(\ref{Ar-an-g1}) replacing $r_b \longleftrightarrow r'_b$ 
and $n_1l_1  \longleftrightarrow n_3l_3$.

The substitution of
$ A_{n_1l_1n_2l_2}^{n_3l_3 n_4l_4,k}({\bf r},{\bf r}',{\rm g}_{\ell})$
to the expression of $O_{22}({\bf r},{\bf r}',g_{\ell})$ which is given
by Eq. (\ref{O22-A}) leads to the analytical expression of the
two-body term of the OBDM, which is of the form
\begin{eqnarray}
O_{22}({\bf r}_b,{\bf r}_b')&=&f_1(r_b,r_b',\cos\omega_{rr'})
\exp\left[-\frac{1+3y}{2(1+y)}r_b^2 -\frac{1}{2}{r_b'}^2\right] 
\nonumber \\
& & + f_1(r_b',r_b,\cos\omega_{rr'})
\exp\left[-\frac{1+3y}{2(1+y)}{r_b'}^2 -\frac{1}{2}r_b^2\right]
\nonumber \\
& & + f_3(r_b,r_b',\cos\omega_{rr'})
\exp\left[-\frac{1+2y}{2}(r_b^2 +{r_b'}^2)\right]
\exp\left[\frac{y^2}{1+2y}({\bf r}_b +{\bf r}_b')^2\right] ,
\end{eqnarray}
where $f_\ell(r_b,r_b',\cos\omega_{rr'}),\  (\ell=1,3)$ are polynomials
of $r_b,\ r_b'$ and $\cos\omega_{rr'}$ which depend also on
$y=\beta b^2$ and the occupation probabilities of the various states.

The corresponding analytical expressions of the matrix elements
$\tilde{A}_{n_1l_1n_2l_2}^{n_3l_3 n_4l_4,k}({\bf k},{\rm g}_\ell)$, 
$(\ell=1,3)$ which contribute to the two-body term of the MD were
found substituting $\phi_{nl}(r)$ with that of the HO wave function into
Eq. (\ref{Ak-O22-1}). The expression of
$ \tilde{A}_{n_1l_1n_2l_2}^{n_3l_3 n_4l_4,k}({\bf k},{\rm g}_1)$, which can
be found easily, has the form
\begin{eqnarray}
\tilde{A}_{n_1l_1n_2l_2}^{n_3l_3n_4l_4,k}({\bf k},{\rm g}_1)&=&
B_0 b^3 (-1)^{n_3} k_b^{2l_3}
\ L_{n_3}^{l_3+\frac{1}{2}}\left(k_b^2 \right) \
\exp \left[- \frac{1+2y}{1+3y} k_b^2 \right]  \nonumber\\
& &\times \sum_{w_1=0}^{n_1}\sum_{w_2=0}^{n_2}\sum_{w_4=0}^{n_4}
\sum_{t=0}^{\frac{1}{2}(l_2+l_4-k)+w_2+w_4}
\tilde{B}_{w_2w_4 t}(y) \left[\frac{1}{2}(l_1-l_3+k)+w_1+t)\right]!  \nonumber\\
& &
\times \frac{(-1)^{w_1}}{w_1!}
\left(  \begin{array}{c}
n_1+l_1+\frac{1}{2} \\
n_1-w_1
\end{array}  \right)
2^{\frac{1}{2}(l_1-l_3)+w_1} \
\ (1+y)^{\frac{1}{2}(l_1+l_3-l_2-l_4)+w_1-w_2-w_4}\nonumber\\
&&
\times (1+3y)^{-\frac{1}{2}(3+l_1+l_3+k)- w_1- t}
\ L_{\frac{1}{2}(l_1-l_3+k)+ w_1+ t}^{l_3+\frac{1}{2}}
\left(\frac{1+y}{2(1+3y)} k_b^2 \right),
\label{Ak-an-g1}
\end{eqnarray}
where
\begin{eqnarray}
\tilde{B}_{w_2 w_4 t}(y)&=& (\sqrt{2}y)^{k+2t}
\left[ \frac{1}{2}(1+l_2+l_4+k) +w_2+w_4 \right]!
\left[ \frac{k-l_2-l_4}{2} -w_2-w_4 \right]_t  \nonumber\\
& &\times \frac{1}{(k+t+\frac{1}{2})!}
\prod_{i=2,4} \frac{(-1)^{w_i+t}}{w_i!t!}
\left( \!\!  \begin{array}{c}
n_{i}+l_{i}+\frac{1}{2} \\
n_{i}-w_i
\end{array} \!\! \right) .
\label{Bk-w2w4}
\end{eqnarray}
The expression of
$ \tilde{A}_{n_1l_1n_2l_2}^{n_3l_3 n_4l_4,k}({\bf k},{\rm g}_3)$
is more complicated. It has the the general form
\begin{eqnarray}
\tilde{A}_{n_1l_1n_2l_2}^{n_3l_3 n_4l_4,k}({\bf k},{\rm g}_3)&=&
\frac{2}{\sqrt{\pi}} B_0 b^3 \exp \left[ -\frac{k_b^2}{1+2y} \right]
\sum_{w_2=0}^{n_2}\sum_{w_4=0}^{n_4}
\sum_{t=0}^{\frac{1}{2}(l_2+l_4-k)+w_2+w_4 }\tilde{B}_{w_2 w_4t}(y)
\nonumber\\
& &\times  (1+2y)^{-\frac{1}{2}(3+l_2+l_4+k)- w_2 - w_4 -t}
I_{n_1l_1}^{n_3l_3,k}(k_b) \ .
\end{eqnarray}

The general expression of the quantity $I_{n_1l_1}^{n_3l_3,k}(k_b)$
is quite complicated. For that reason we calculated it for various cases
which are needed for the $s$-$p$ and $s$-$d$ shell nuclei. The various cases and
the corresponding expressions of $I_{n_1l_1}^{n_3l_3,k}(k_b)$ are given
bellow.

\subsubsection{ Case 1: $n_1l_1=n_3l_3$ and $k=0$}
               %%%%%%%%%%%%%%%%%%%%%%%%%%%%%%%%%%%%%
\begin{eqnarray}
I_{n_1l_1}^{n_1l_1,0}(k_b)&=&
\sum_{\rho=0}^{[l_1/2]} \sum_{\sigma=0}^{n_1} \sum_{\tau=0}^{\rho+\sigma}
\sum_{\nu=0}^{2(\rho+\sigma-\tau)} \sum_{\alpha=0}^{l_1-2\rho}
\sum_{\mu=0}^{n_1-\sigma} \sum_{w_1=0}^{n_1-\sigma-\mu}
\sum_{w_3=0}^{\mu}
\frac{(-1)^{\rho+\alpha+ \tau +w_1+w_3}}{n_1! \rho ! \sigma ! w_1! w_3!}
L_{\alpha+w_1+\tau+\nu}^{\frac{1}{2}}
\left(\frac{k_b^2 }{1+2y} \right)     
\nonumber \\
& &\times 
\left(\!\!  \begin{array}{c}
l_1 - 2\rho \\
\alpha \end{array}  \!\! \right)
\left(\!\!  \begin{array}{c}
\rho +\sigma \\
\tau \end{array}\!\!  \right)
\left(\!\!  \begin{array}{c}
2(\rho + \sigma -\tau) \\
\nu \end{array}\!\!  \right)
\left(\!\!  \begin{array}{c}
n_1 + \sigma -\mu \\
n_1 -\sigma -\mu -w_1 \end{array}\!\!  \right)
\left(\!\!  \begin{array}{c}
\mu  +l_1 - \frac{1}{2} \\
\mu -w_3 \end{array}\!\!  \right)       \nonumber\\
& & \nonumber\\
& &
\times \frac{(\alpha +w_1+\tau+\nu)!}{2\tau+1}
\frac{(2l_1 -2\rho)! (n_1+l_1+\frac{1}{2} )!
(l_1+w_3+t+2\sigma-\alpha-\tau-\nu+\frac{1}{2} )!}
{(l_1-\rho)! (l_1 - 2\rho)!(\sigma+l_1+\frac{1}{2})!}    \nonumber\\
& &\nonumber\\
& &\times 
2^{-l_1+w_1+w_3+t+2\tau} (1+2y)^{l_1-w_1+w_3+t + 2(\sigma- \alpha- \tau- \nu)}
(1+4y)^{-\frac{3}{2}-l_1-w_3-t-2\sigma+\alpha+\tau+\nu} 
\ .
\label{nl1nl3-1}
\end{eqnarray}

\subsubsection{ Case 2: $n_1l_1=n_3l_3=01$ and $k=2$ }
                        %%%%%%%%%%%%%%%%%%%%%%%%%%%%%%
\begin{eqnarray}
I_{01}^{01,2}(k_b)&=&
2^{1+t} \left( \frac{5}{2}+t \right)! (1+2y)^{2+t} (1+4y)^{-\frac{7}{2}-t} \ .
\label{nl1nl3-5}
\end{eqnarray}

\subsubsection{ Case 3: $n_1l_1=n_3l_3=02$ and $k=2,4$}
                        %%%%%%%%%%%%%%%%%%%%%%%%%%%%%%
\begin{eqnarray}
I_{02}^{02,k}(k_b)&=&
2^{\frac{k}{2}+t}(1+2y)^{1+t} (1+4y)^{-\frac{7}{2}-t}
\nonumber\\
& &\times \left[ \left(\frac{5+k}{2} + t\right)! (1+2y)^{1+\frac{k}{2}}
(1+4y)^{-\frac{k}{2}} - \delta_{2k}\frac{7}{3} \left(\frac{5}{2}+t\right)! 
L_1^{\frac{1}{2}}\left(\frac{k_b^2}{1+2y}\right) \right] .
\label{nl1nl3-4}
\end{eqnarray}

\subsubsection{ Case 4: $n_1l_1 \ne n_3l_3$ and $l_1=0$ or/and $l_3=0$}
                %%%%%%%%%%%%%%%%%%%%%%%%%%%%%%%%%%%%%%%%%%%%%%%%%%%%%%
\begin{eqnarray}
I_{n_1l_1}^{n_3l_3,k}(k_b)&=&
\sum_{w_1=0}^{n_1} \sum_{w_3=0}^{n_3}
\sum_{\rho_1=0}^{[l_1/2]} \sum_{\rho_3=0}^{[l_3/2]}
\sum_{\tau_1=0}^{l_1-2\rho_1} \sum_{\tau_3=0}^{l_3-2\rho_3}
\sum_{\sigma_1=0}^{\rho_1+w_1} \sum_{\sigma_3=0}^{\rho_3+w_3}
\sum_{\nu=0}^{\rho_1+\rho_3+w_1+w_3-\sigma_1-\sigma_3}
\frac{1+(-1)^{\tau_1+\tau_3+\sigma_1+\sigma_3}}
{2(\tau_1+\tau_3+\sigma_1+\sigma_3 +1)} 
\nonumber \\
& &
\times \left[ \prod_{i=1,3} (-1)^{\rho_i +w_i+ \tau_3 +\sigma_3}
\frac{2^{\frac{1}{2}k-l_i+\sigma_i+t}(2 l_i -2\rho_i)!}
{w_i! (l_i-\rho_i)!(l_i-2\rho_i)!}
\left(\!\!  \begin{array}{c}
n_i+l_i +\frac{1}{2} \\
n_i -w_i \end{array}\!\!  \right)
\left(\!\!  \begin{array}{c}
l_i-2\rho_i \\
\tau_i \end{array} \!\!  \right)
\left(\!\!  \begin{array}{c}
\rho_i+w_i \\
\sigma_i \end{array}\!\!  \right) \right.  \nonumber \\
& &
\times \left. (1+2y)^{\frac{1}{2}(k+l_i)+w_i-\sigma_i-\tau_i-2\nu+t}
(1+4y)^{-\frac{1}{2}(3+k+l_i-\sigma_i-\tau_i)- w_i -t + \nu} 
\frac{  }{  }  \right]
\nonumber \\
& &
\times  \left[\frac{1+k+l_1+l_3-\sigma_1-\sigma_3-\tau_1-\tau_3}{2}
+t+w_1+w_3-\nu \right]!
\left[\frac{\sigma_1+\sigma_3+\tau_1+\tau_3}{2}+\nu \right]! 
\nonumber \\
& &
\times \left(\!\!  \begin{array}{c}
\rho_1+\rho_3+w_1+ w_3-\sigma_1-\sigma_3 \\
\nu \end{array} \!\! \right)
L_{\frac{1}{2}(\sigma_1+\sigma_3+\tau_1+\tau_3+2\nu)}^{\frac{1}{2}}
\left(\frac{k_b^2 }{1+2y}  \right)  .
\label{nl1nl3-2}
\end{eqnarray}

\subsubsection{ Case 5: $n_1l_1=01,\ \ n_3l_3=02$ or
                        $n_1l_1=02,\ \ n_3l_3=01$, \ and \ $k=1,3$}
                        %%%%%%%%%%%%%%%%%%%%%%%%%
\begin{eqnarray}
I_{01}^{02,k}(k_b)&=& I_{02}^{01,k}(k_b)\ =\
2^{\frac{k}{2}+t} \left(1+\frac{k}{2}+t \right)!
(1+2y)^{\frac{1}{2}(3+k)+t} (1+4y)^{-3-\frac{k}{2}-t}    \nonumber\\
& &
 \times \left[ \left(2+\frac{k}{2}+t \right) +
 \left(\frac{4}{k^2+k-6}-\frac{2}{3}\right)\frac{1+4y}{(1+2y)^2}
 L_1^{\frac{1}{2}}\left(\frac{k_b^2}{1+2y}\right) \right] .
\label{nl1nl3-3}
\end{eqnarray}

The substitution of
$ \tilde{A}_{n_1l_1n_2l_2}^{n_3l_3 n_4l_4,k}({\bf k}, {\rm g}_{\ell})$
to the expression of $\tilde{O}_{22}({\bf k},{\rm g}_{\ell})$ which is given
by Eq. (\ref{O22-A}) leads to the analytical expression of the
two-body term of the MD, which is of the form
\begin{eqnarray}
\tilde{O}_{22}(k)&=&
{\tilde f}_1(k_b^2)\exp\left[-\frac{1+2y}{1+3y} k_b^2 \right] +
{\tilde f}_3(k_b^2)\exp\left[-\frac{1}{1+2y} k_b^2 \right] ,
\end{eqnarray}
where $\tilde{f}_\ell(k_b^2), \ (\ell=1,3)$ are polynomials of $k_b^2$ 
which depend also on $y=\beta b^2$ and the occupation probabilities of 
the various states. Similar expressions have been found for the mean 
value of the kinetic energy.

It should be noted that, although the above expressions of the matrix
elements $A$ and $\tilde{A}$  seem to be quite complicated, they can easily
be used for analytical calculations with programs such as Macsyma or
Mathematica. As the above expressions have been found for the ($N=Z$)  
$s$-$p$ and $s$-$d$ shell nuclei they can be used for the systematic study 
of the OBDM and MD in this region of nuclei.

%and the terms $O_{22}({\bf r},{\bf r}')$
%and $\tilde{O}_{22}(k)$

\section{RESULTS AND DISCUSSION}
         %%%%%%%%%%%%%%%%%%%%%%

The calculations of the MD for the various $s$-$p$ and $s$-$d$ shell nuclei,
with $N=Z$, have been carried out on the basis of Eq. (\ref{ncor-1}) and the
analytical expressions of the one- and two-body terms which were given in
Sec. III. Two cases have been examined, named case 1 and case 2
corresponding to the analytical calculations with HO orbitals without
and with SRC, respectively.

The parameters $b$ and $\beta$ of the model in case 1 and
for $^4$He, $^{16}$O, $^{36}$Ar and $^{40}$Ca in case 2 were the ones which
have been determined in our previous work \cite{Massen99} by fit of the
theoretical $F_{ch}(q)$, derived with the same cluster expansion, to the 
experimental one. These values of the parameters are given in Table I.
The values of the correlation parameter $\beta$ of the open shell nuclei 
which have been reported in Ref. \cite{Massen99} were quite large. 
That is the correlations for these nuclei were quite small.
The MD of the open shell nuclei, which we found with these values of the
parameters, had a high momentum tail at  values of $k$ larger than 
expected. As that seems to us quite unreasonable we tried to redetermine
more carefully the parameters of the model by fit of the theoretical
$F_{ch}(q)$ to the experimental one in order to obtain a better fit.

The new values of $b$ and $\beta$ for case 2 and for
$^{12}$C, $^{24}$Mg, $^{28}$Si and $^{32}$S are shown in Table I.
The  theoretical $F_{ch}(q)$ for these nuclei, which are shown in Fig. 1,
are closer to the experimental data than they were in Ref. \cite{Massen99}.
From the values of $\chi^2$, which have been found in cases 1 and 2 and
also from Fig. 1 it can been seen that the inclusion of SRC's improves
the fit of the form factor of the above mentioned nuclei.
Also, all the diffraction minima, even the third one which  seems
to exist in the experimental data of $^{24}$Mg, $^{28}$Si and $^{32}$S
are reproduced in the correct place.

Although the values of the parameters $b$ and $\beta$, for the open shell
nuclei, are different from those reported in Ref. \cite{Massen99},
their behaviour, still, indicates that there should be a shell effect
in the case of closed shell nuclei.
This behaviour has an effect on the MD of nuclei as it is seen
from Fig. 2, where the MD, of the various $s$-$p$ and $s$-$d$ shell nuclei
calculated with the values of $b$ and $\beta$ of Table I for case 2,
have been plotted.
It is seen that the inclusion of SRC's increases considerably the high 
momentum component of $n(k)$, for all nuclei we have considered.
Also, while the general structure of the high momentum component of
the MD for $A=4,\ 12,\ 16,\ 24,\ 28,\ 32, 36,\ 40$, is almost the same, 
in agreement with other studies \cite{Antonov,Zabolitzky78,Traini85,Ciofi84},
there is an $A$ dependence of $n(k)$ both at small values of
$k$ and in the region $2\ {\rm fm}^{-1} < k < 5\ {\rm fm}^{-1}$.
The $A$ dependence of the high momentum component of $n(k)$ is larger in 
the open shell nuclei than in the closed shell nuclei. It is seen that 
the high momentum component is almost the same for the closed shell nuclei
$^4$He, $^{16}$O and $^{40}$Ca as expected from other studies
\cite{Antonov,Zabolitzky78,Ciofi84}.

In the previous analysis, the nuclei $^{24}$Mg, $^{28}$Si and $^{32}$S
were treated as $1d$ shell nuclei, that is, the occupation probability
of the $2s$ state was taken to be zero. The formalism of the
present work has the advantage that the occupation probabilities
of the various states can be treated as free parameters in the
fitting procedure of $F_{ch}(q)$. Thus, the analysis can be made with
more free parameters.
For that reason we considered case $2^*$ in which the occupation 
probability $\eta_{2s}$ of the nuclei $^{24}$Mg, $^{28}$Si and $^{32}$S 
was taken to be a free parameter together with the parameters $b$ and 
$\beta$. We found that the $\chi^2$ values become better, compared to 
those of case 2 and the $A$ dependence of the parameter $\beta$ is not 
so large as it was before. The new values of $b$ and $\beta$ are shown 
in Table I and the theoretical $F_{ch}(q)$ in Fig 1.
The values of the occupation probability  $\eta _{2s}$ of the
above-mentioned three nuclei are 0.19982, 0.17988 and 0.50921,
respectively, while the corresponding values of $\eta_{1d}$, which
can be found from the values of $\eta_{2s}$ through the relation
\[ \eta_{1d} = [(Z-8) - 2 \eta_{2s}]/10,
\]
are 0.36004,  0.56402 and 0.69816, respectively.
The MD of these three nuclei together with the closed shell nuclei
$^4$He, $^{16}$O and $^{40}$Ca found in case 2 are shown in Fig. 3.
It is seen that the $A$ dependence of the high momentum component
is now not so large as it was in case 2.
As $F_{ch}(q)$ calculated in case $2^*$ is closer to the experimental data
than in case 2, we might say that this result is in the correct direction, 
that is the high momentum component of the MD of nuclei is almost the same. 
We would like to mention that experimental data for $n(k)$ are not directly
measured but are obtained by means of $y$-scaling analysis \cite{Ciofi91}
and only for $^4$He and $^{12}$C in $s$-$p$ and $s$-$d$ shell region.
We expect that the above conclusion could be corroborated if new
experimental data are obtained in the future for  MD for several nuclei and 
we carry out a simultaneous fit both to MD and to form factors.

Finally, in table I we give the one and the two-body terms of the mean
kinetic energy, $\langle {\bf T} \rangle$, of the various $s$-$p$ and
$s$-$d$ shell nuclei calculated on the basis of Eq. (\ref{Kin-1}), as well as
the rms charge radii, $\langle r^2_{ch} \rangle^{1/2}$ which are compared
with the experimental values.
It is seen that the introduction of SRC's (in case 2) increases the
mean kinetic energy relative to case 1
($(\langle{\bf T}_{case2}\rangle -\langle{\bf T}_{case1}\rangle)/
\langle{\bf T}_{case2}\rangle$) about $50\%$ in $^4$He and $23\%$ in 
$^{24}$Mg. This relative increase follows the fluctuation of the parameter $
\beta$. Also the values of the kinetic energy in percents,
$100 \langle{\bf T}_{SRC}\rangle/\langle{\bf T}_{Total}\rangle$,
as well as the ratio $<{\bf T}_{Total}>/\langle{\bf T}_{HO}\rangle$
follow the fluctuation of the parameter $\beta$.
In closed shell nuclei there is an increase of the above values
by the increasing of mass number.

%======================

\section{SUMMARY}
        %%%%%%%%
In the present work, general expressions for the correlated OBDM
and MD have been found using the factor cluster
expansion of Clark and co-workers. These expressions can be used for
analytical calculations, with HO orbitals and in principle for numerical
calculations with more realistic orbitals.

The analytical expressions of the OBDM, MD and mean kinetic energy
for the $s$-$p$ and $s$-$d$ shell nuclei, which have been found, are functions
of the HO parameter $b$, the correlation parameter $\beta$ and the
occupation probabilities of the various states.
These expressions are suitable for the systematic study of the above
quantities for the $N=Z$, $s$-$p$ and $s$-$d$ shell nuclei and also for the
study of the dependence of these quantities on the various parameters.

It is found that, while the general structure of the MD at high momenta is
almost the same for all the nuclei we have considered, in agreement with
other studies, there is an $A$ dependence on $n(k)$ both at small values
of $k$ and the high momentum component. The $A$ dependence of the high 
momentum component becomes quite small if the occupation probability of the
$2s$-state for the nuclei $^{24}$Mg, $^{28}$Si and $^{32}$S is treated
as a free parameter in the fitting procedure of the charge form factor.

\section*{ }
%ACKNOWLEDGMENTS}

The authors would like to thank Professor M.E. Grypeos and Dr. C.P. Panos
for useful comments on the manuscript. One of the author (Ch.C.M) would
like to thank Dr. P. Porfyriadis for technical assistants.

%\newpage
%%%%%%%%%%%%%%%%%%%

\newpage

\begin{table}
\caption{The values of the parameters $b$ and $\beta$, of the mean kinetic 
energy per nucleon, $\langle {\bf T} \rangle$ and of the rms charge radii,
$\langle r_{ch}^{2}\rangle^{1/2}$, for various $s$-$p$ and $s$-$d$ shell 
nuclei, determined by fit to the experimental $F_{ch}(q)$. Case 1 refers to 
the  HO wave function without SRC and case 2 when SRC are included. Case 2* 
is the same as case 2 but with the occupation probability of the state $2s$ 
taken to be as a free parameter. The experimental rms charge radii are from 
Ref. [41].}
%\cite{DeVries82} }
\begin{center}
\begin{tabular}{c c c c c c c c c}
\hline
 & & & & & & & &   \\
Case& Nucleus & $b$ [fm] & $\beta$ [fm$^{-2}$] &
\multicolumn{3}{c}{$\langle {\bf T} \rangle$ [Mev]} &
\multicolumn{2}{c}{$\langle r_{ch}^{2}\rangle^{1/2}$ [fm]} \\
%\cline{5-9}
 & & & & & & & & \\
 & & & &HO&SRC&Total&Theor. & Expt.\\
&&&&&&&& \\
\hline
&&&&&&&& \\
 1 & $^{4}$He & 1.4320 & --     & 15.166  &  --   &15.166  & 1.7651  & 1.676(8)   \\
 2 & $^{4}$He & 1.1732 & 2.3126 & 22.594 & 7.310 &29.904 &  1.6234 &             \\
&&&&&&&& \\
 1 & $^{12}$C & 1.6251 & --    & 17.010  & --    &17.010 &2.4901   &2.471(6)\\
 2 & $^{12}$C & 1.5190 &2.7468 & 19.469  &6.111  &25.580 &2.4261  &   \\
&&&&&&&& \\
 1& $^{16}$O & 1.7610 & --     & 15.044   &---   &15.044 &2.7377   & 2.730(25)  \\
 2& $^{16}$O & 1.6507 & 2.4747 & 17.121   &6.493 &23.614 &2.6802  &      \\
&&&&&&&& \\
 1 & $^{24}$Mg    & 1.8495 & --     & 16.162 &--   &16.162 &3.1170 & 3.075(15)  \\
 2 & $^{24}$Mg    & 1.8103 & 4.2275 & 16.870 &4.239&21.109 & 3.0948 &   \\
 2*& $^{24}$Mg & 1.7473 & 2.4992 & 18.109 &6.505&24.614 &3.0638 &  \\
&&&&&&&& \\
 1 & $^{28}$Si   & 1.8941 & --    &16.099    &--   &16.099  &3.2570 &  3.086(18) \\
 2 & $^{28}$Si   & 1.8236 &3.0020 &17.369    &5.564&22.933  &3.2159 &      \\
 2*&$^{28}$Si & 1.7774 &2.4440 &18.283    &6.922&25.205  &3.1835 &       \\
&&&&&&&& \\
 1 & $^{32}$S     & 2.0016 & --    & 14.878 &--  &14.878    &3.4830 & 3.248(11) \\
 2 & $^{32}$S     & 1.9368 &3.0659 & 15.891 &4.976&20.867   &3.4425 &       \\
 2*& $^{32}$S  & 1.8121 &2.6398 & 18.154 &6.761&24.915   &3.2822 &    \\
&&&&&&&& \\
 1 & $^{36}$Ar & 1.8800 & --    & 17.273 &--  &17.273 &3.3270 &3.327(15)  \\
 2 & $^{36}$Ar & 1.8007 &2.2937 & 18.827 &8.590&27.417 &3.3343&           \\
&&&&&&&& \\
 1 & $^{40}$Ca & 1.9453 & --    & 16.437 &--& 16.437   &3.4668 &3.479(3)  \\
 2 & $^{40}$Ca & 1.8660 &2.1127 & 17.863 &8.754&26.617 &3.5156 &    \\
&&&&&&&& \\
\hline
\end{tabular}
\end{center}
\end{table}

%%%%%%%%%%%%%%%%%%%%%%%%%%%%%%%%%%%%%%%%%%%%%%%%%%%%%%%%
\newpage

\begin{figure}
\label{ff1-fig}
\begin{center}
\begin{tabular}{cc}
{\epsfig{figure=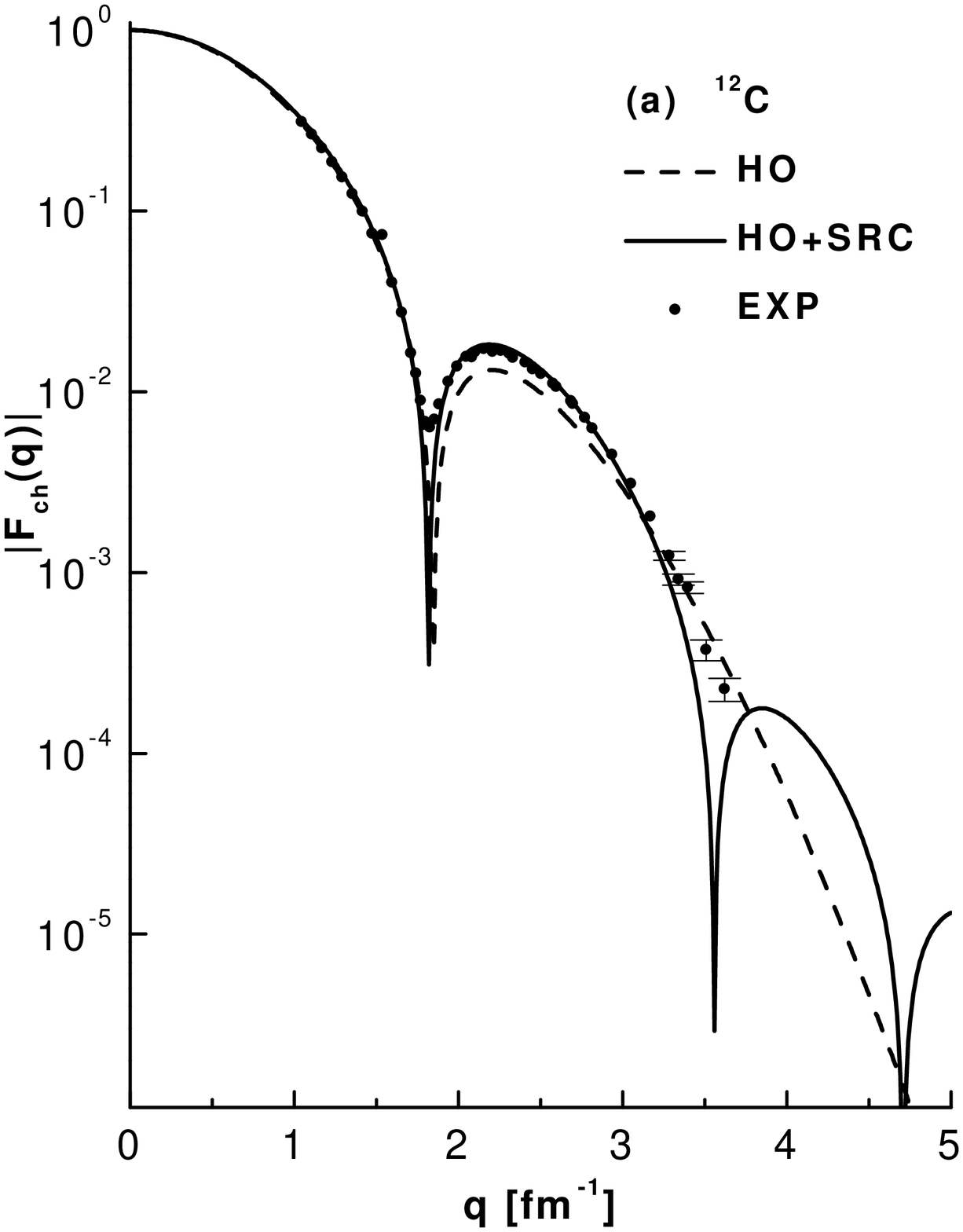,width=6.cm} } &
{\epsfig{figure=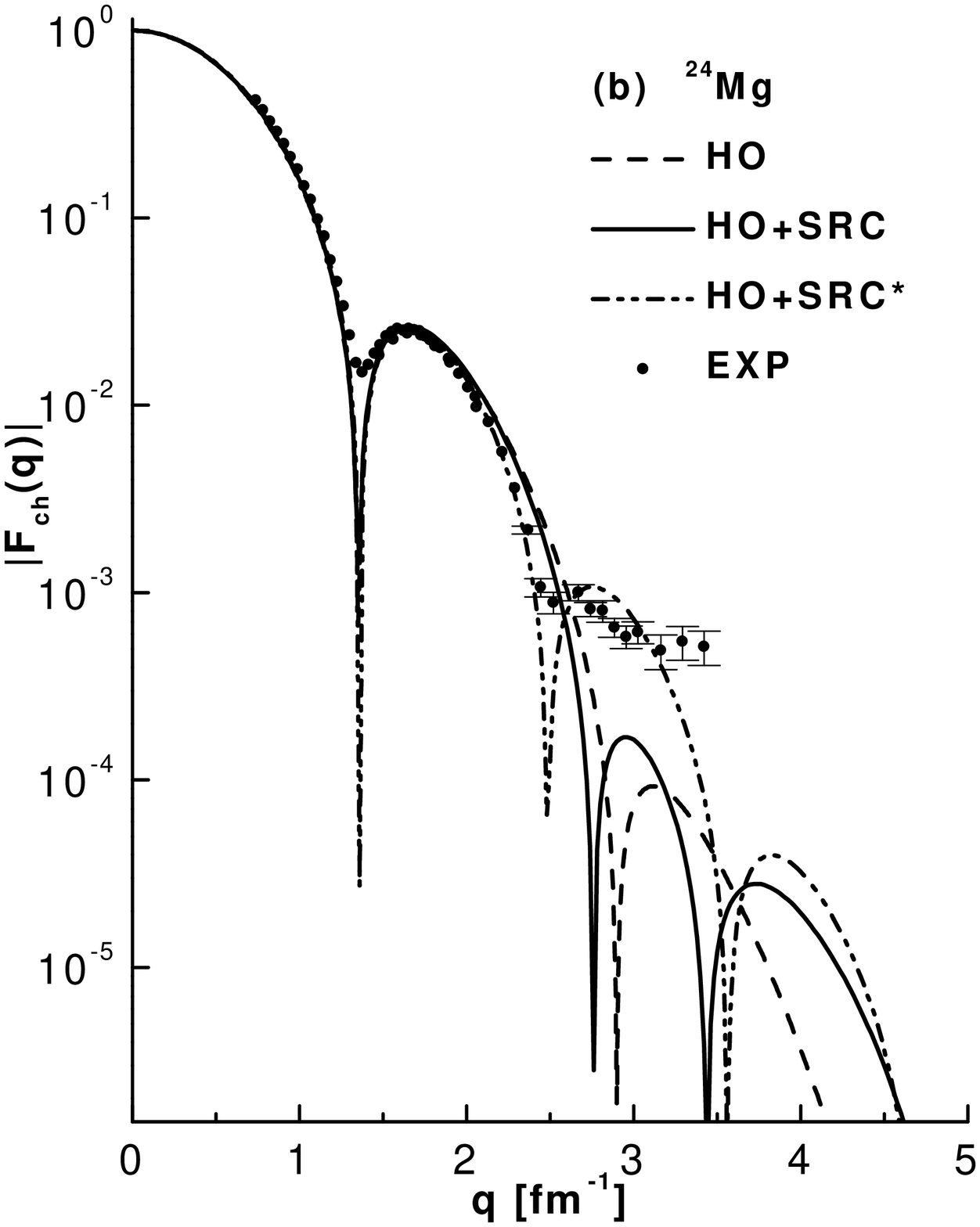,width=6.cm} }  \\
 & \\
 & \\
{\epsfig{figure=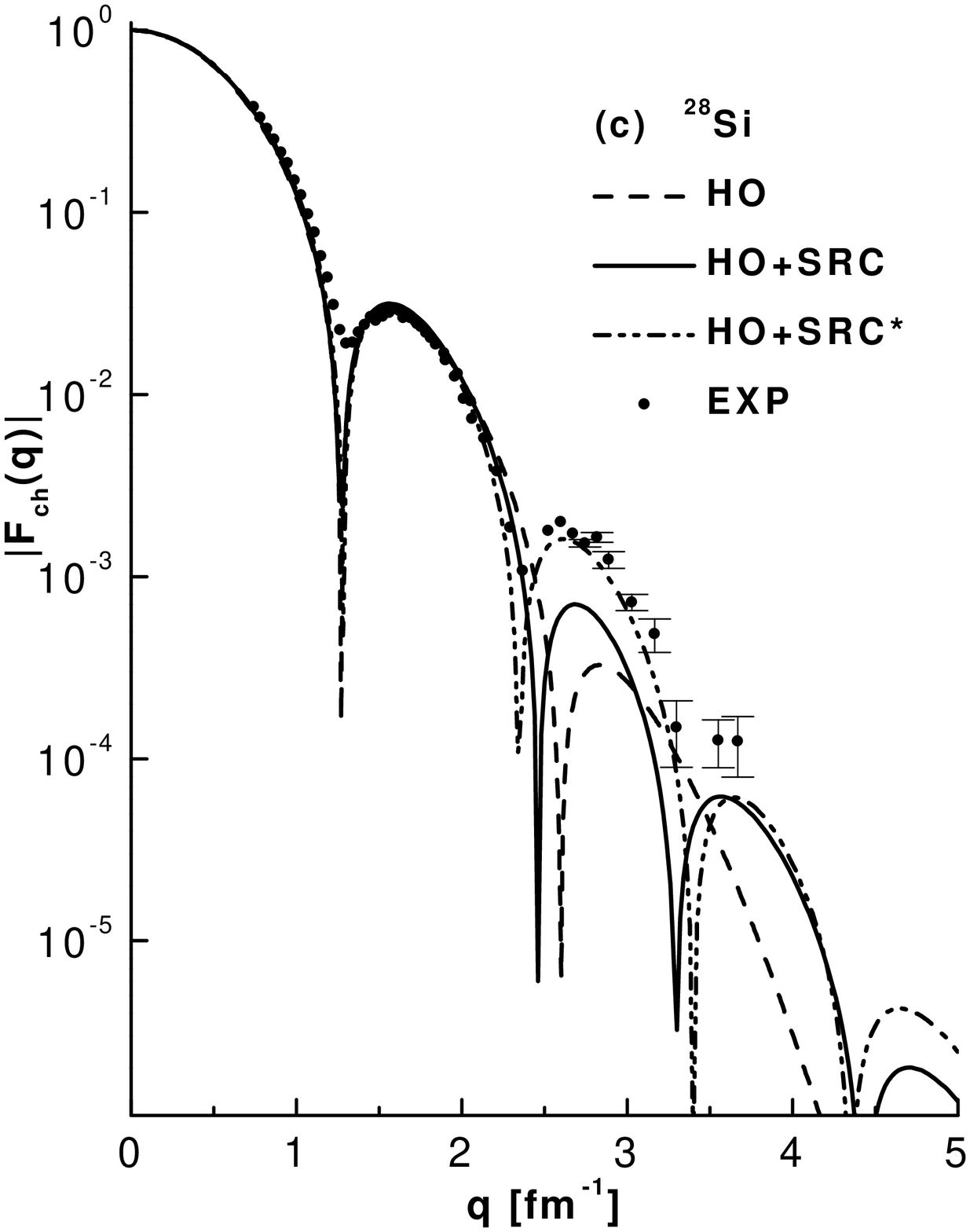,width=6.cm} } &
{\epsfig{figure=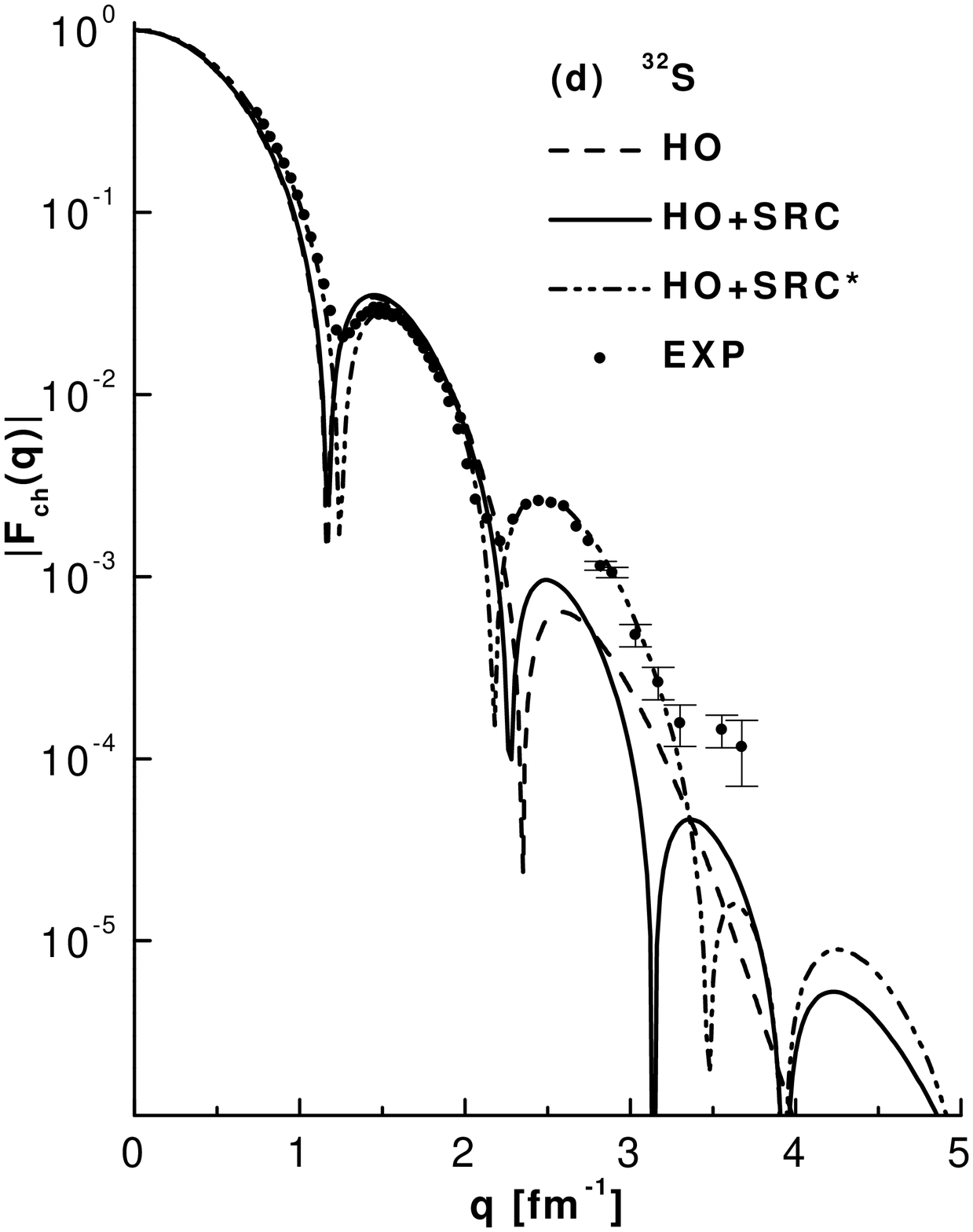,width=6.cm} }  \\
\end{tabular}
\end{center}
\caption{The charge form factor of the nuclei: $^{12}$C (a), $^{24}$Mg (b),
$^{28}$Si (c) and $^{32}$S (d) for various cases. Case HO+SRC* corresponds 
to the case when the occupation probability $\eta_{2s}$ is treated as a 
free parameter. The experimental points of $^{12}$C are from Ref. [42]
%\cite{Sick70}
and for the other nuclei from Ref. [43].}
%\cite{Li74}. }
\end{figure}

%%%%%%%%%%%%%%%%%%%%%%%%figure 2

\begin{figure}
\label{momn-fig}
\centerline{\epsfig{figure=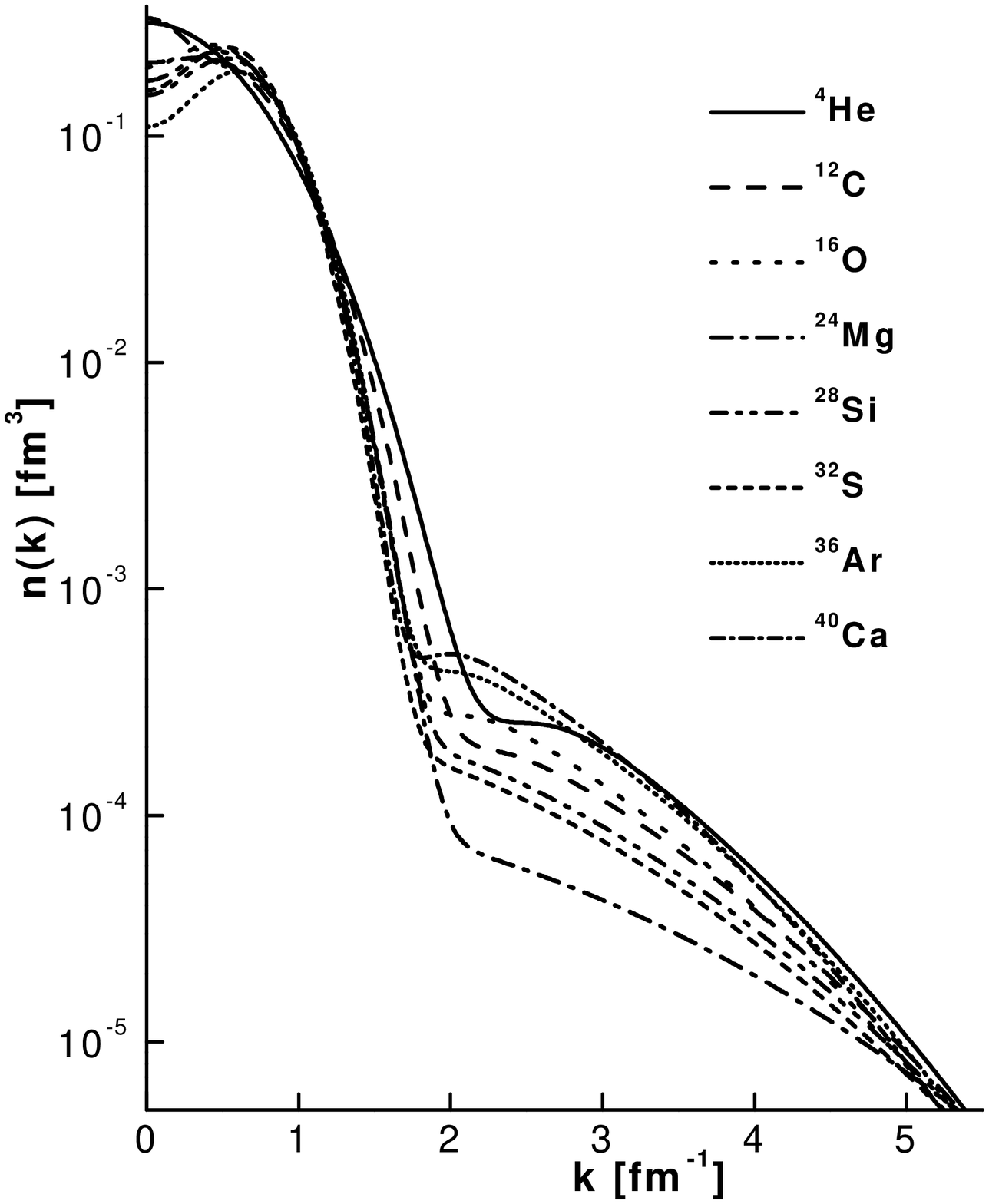,width=7cm} }
\caption{The correlated momentum distribution for various  $s$-$p$ and $s$-$d$ 
shell nuclei calculated with the parameters $b$ and $\beta$ of case 2 when 
the nuclei $^{24}$Mg, $^{28}$Si, $^{32}$S and $^{36}$Ar were treated as 
$1d$ shell nuclei. The normalization is $\int n({\bf k}) {\rm d} {\bf k}=1$.}
\end{figure}

\vspace*{1cm}
%%%%%%%%%%%%%%%%%%%%%%%%figure 3
\begin{figure}
\label{mom2s-fig}
\centerline{\epsfig{figure=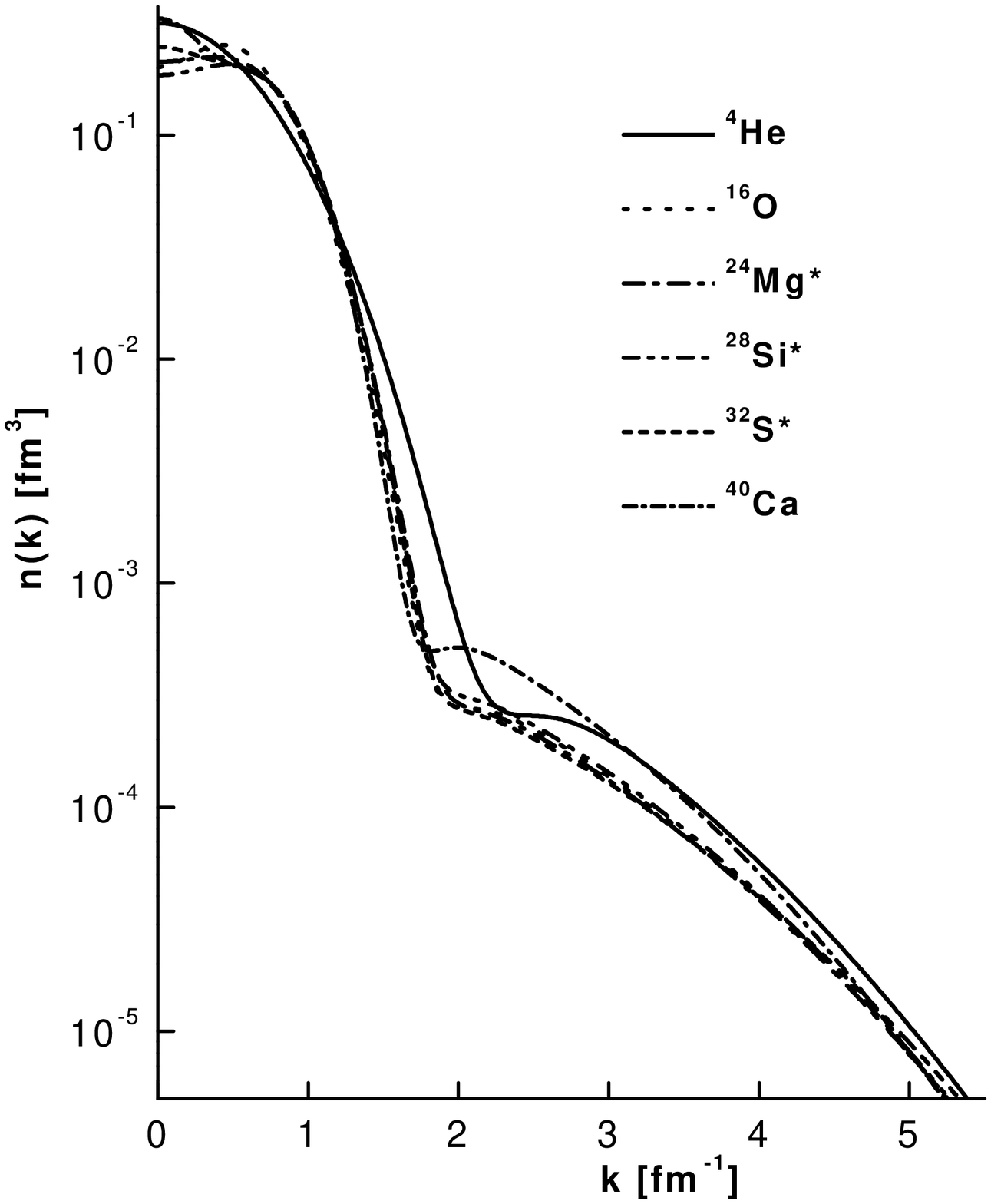,width=7cm} }
\caption{The correlated momentum distribution for the closed shell nuclei
$^{4}$He, $^{16}$O, $^{40}$Ca calculated as in Fig. 2 and for the open shell 
nuclei $^{24}$Mg, $^{28}$Si, $^{32}$S calculated with the parameters $b$, 
$\beta$ and $\eta_{2s}$ of case 2* when they were treated as $1d$-$2s$ 
shell nuclei. The normalization is as in Fig. 2.}
\end{figure}

\end{document}